\documentclass[pss]{wiley2sp} 
\usepackage{amsmath}
\usepackage{amssymb}

\newcommand{\bra}[1]{\langle{#1}|}
\newcommand{\ket}[1]{|{#1}\rangle}

\begin{document}

\title{Nuclear spins in nanostructures}

\titlerunning{Nuclear spins in nanostructures}

\author{%
  W. A. Coish\textsuperscript{\textsf{\bfseries 1,\Ast}},
  J. Baugh\textsuperscript{\textsf{\bfseries 1,2}}}

\authorrunning{W. A. Coish and J. Baugh}

\mail{e-mail
  \textsf{wcoish@iqc.ca}}

\institute{%
  \textsuperscript{1}\,Institute for Quantum Computing and Department of Physics and Astronomy, University of Waterloo, 200 University Ave. W., Waterloo, ON, Canada\\
  \textsuperscript{2}\,Department of Chemistry, University of Waterloo, 200 University Ave. W., Waterloo, ON, Canada\\
}

\received{XXXX, revised XXXX, accepted XXXX} 
\published{XXXX} 


\abstract{%
%
%
%
\abstcol{%
  We review recent theoretical and experimental advances toward understanding the effects of nuclear spins in confined nanostructures.  These systems, which include quantum dots, defect centers, and molecular magnets, are particularly interesting for their importance in quantum information processing devices, which aim to coherently manipulate single electron spins with high precision.   On one hand, interactions between confined electron spins and a nuclear-spin environment provide a decoherence source for the electron, and on the other, a strong effective magnetic field that can be used to execute local coherent rotations.  A great deal of effort has been directed toward understanding the details of the relevant decoherence processes and to find new methods to manipulate the coupled electron-nuclear system.

A sequence of spectacular new results have provided understanding of spin-bath decoherence, nuclear spin diffusion, and preparation of the nuclear state through dynamic polarization and more general manipulation of the nuclear-spin density matrix through ``state narrowing''.  These results demonstrate the richness of this physical system and promise many new mysteries for the future.}{}}

%
%

\titlefigure[height=6cm,width=7.5cm]{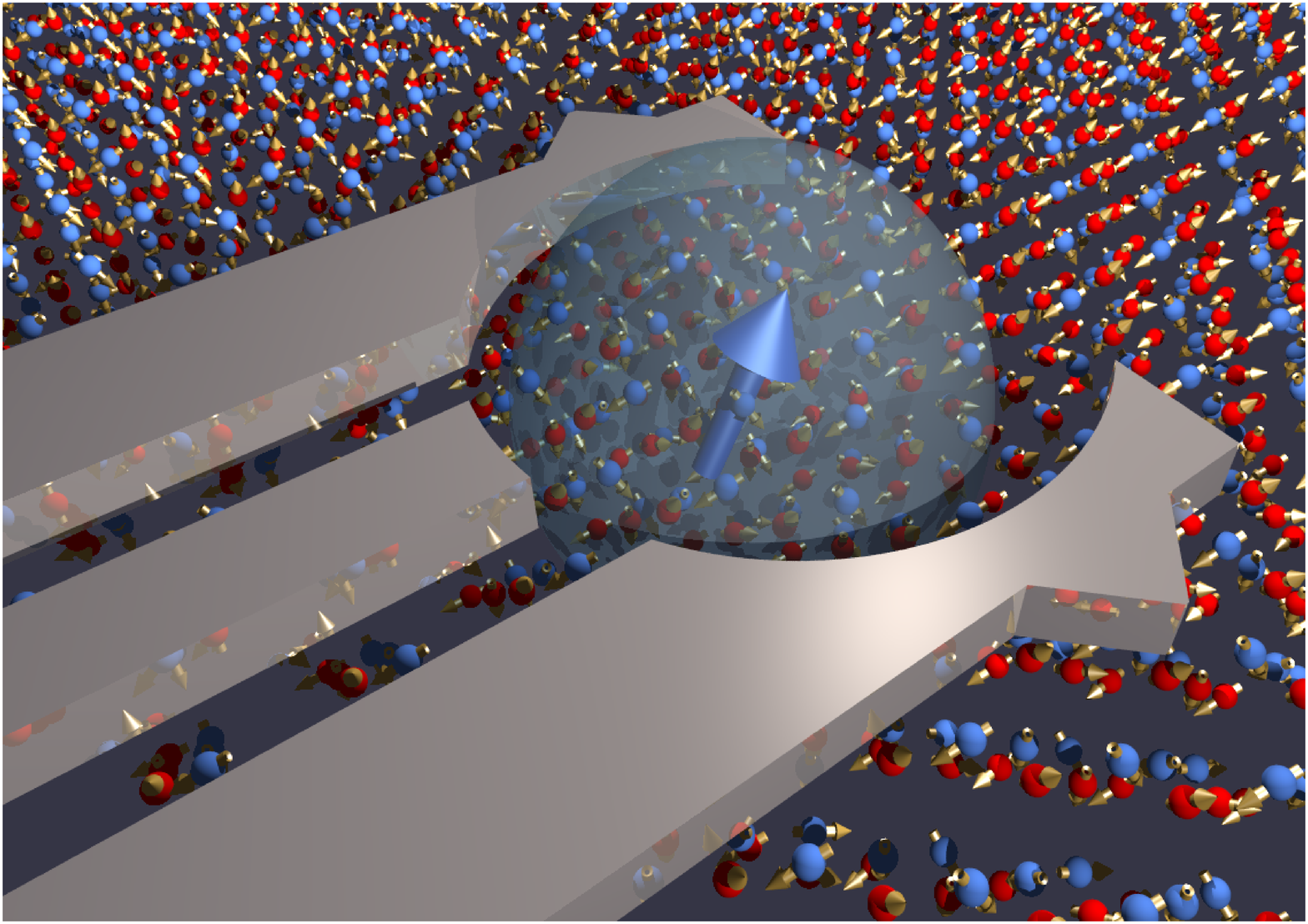}

\titlefigurecaption{%
  Illustration of an electron confined to a gated lateral quantum dot.  The electron envelope function is represented by a translucent blue sphere.  An electron spin (large blue arrow) interacts with many nuclei at atomic sites (red and blue spheres), which typically carry a finite spin (represented by small yellow arrows).}

\maketitle   

\section{Introduction}
The last several years have seen a series of breakthroughs in single-spin measurement and manipulation, motivated in large part by the potential for future quantum information processing devices \cite{Loss1998a,Kane1998a}.  The spin coherence times for confined electrons in semiconductor quantum dots \cite{Burkard1999a,Khaetskii2002a,Merkulov2002a,DeSousa2003a,Coish2004a,Petta2005b,Greilich2006a,Koppens2008a}, phosphorus donor impurities in silicon \cite{Tyryshkin2003a,Abe2004a}, nitrogen vacancy (NV) centers in diamond \cite{Jelezko2004a,Gaebel2006a,Childress2006a}, and in molecular magnets \cite{Ardavan2007a,Troiani2008a} is typically limited by the interaction between the electron and nuclear spins in the host material.  The coherent manipulation of electron spins therefore requires a complete understanding of the \emph{nuclear} spins in these materials, typically in the presence of localized electrons.

A great deal of work has been done many years ago on ensembles of electron spins at donor impurities, including experimental \cite{Honig1954a,Feher1959a,Mims1961a} and theoretical \cite{Pines1957a,Klauder1962a} studies of electron spin relaxation \cite{Pines1957a,Feher1959a} and dephasing \cite{Mims1961a}, dynamical nuclear polarization \cite{Honig1954a,Abragam1958a,Paget1981a}, and nuclear spin diffusion \cite{Paget1981a}.  Much can be learned (and has been learned) from these past studies, but at the same time, new experiments performed on \emph{single} isolated spins in solids provide a new system, which cannot be generically described by previous work relying on inhomogeneous ensembles.

To avoid complications due to nuclear spins, it may be advantageous to construct nanostructures from graphene \cite{Trauzettel2007a}, carbon nanotubes \cite{Mason2004a,Graeber2006a}, or Si/SiGe \cite{Shaji2008a}, where the majority isotopes carry no nuclear spin.  However, in addition to detrimental effects of nuclear spins (decoherence), a polarized nuclear-spin system provides an effective magnetic field, which can be used to split electron-spin states, allowing for highly local control of single spins \cite{Coish2007a}.  Alternatively, long-lived nuclear spin states may serve as a robust quantum \cite{Taylor2003a,Morton2008a}, or classical \cite{Austing2009a} memory device.  Newfound understanding in methods of generating large sustained nuclear polarization, coupled with knowledge of the dissipation of this polarization may help significantly advance the field of biomedical imaging using nuclear magnetic resonance (NMR), which often relies on a large nuclear polarization to enhance sensitivity \cite{Aptekar2009a}.  Finally, a careful analysis of electron-nuclear interactions as well as dynamic polarization in nanostructures can reveal new insights in strongly-correlated electron systems, through the use of innovative NMR techniques \cite{Dobers1988a,Smet2002a,Desrat2002a,Stern2004a,Dean2008a,Kawamura2009a,Kempf2008a,Hirayama2009a}.

In this article, we do not set out to describe all of the interesting physical effects involving nuclear spins in nanostructures.  Instead, we attempt to give a brief review of the most important fundamental concepts needed to understand the relevant phenomena and summarize what we feel to be some of the most important recent results from the field.

The rest of this article is organized as follows:  In Section \ref{sec:Interactions} we review the major sources of interaction for nuclear spins in a solid, with a focus on nanostructures, allowing for the possibility of a strongly-interacting many-electron system in the nuclear environment.  Section \ref{sec:DNP} gives a summary of recent results on dynamic nuclear polarization (DNP) for nuclear spins in quantum dots.  In Section \ref{sec:decoherence} we review the important problem of decoherence for a single electron spin interacting with a bath of nuclear spins, and in Section \ref{sec:conclusions} we conclude with an overview of what we believe to be some of the outstanding questions in this emerging field.

\section{Nuclear spin interactions}\label{sec:Interactions}

Before moving on to a survey of the recent literature, here we review the relevant Hamiltonians for nuclear spins in a solid.  A detailed discussion of these interactions can be found, for instance, in the well-known books by Abragam \cite{Abragam1962a} and Slichter \cite{Slichter1980a}, but here we focus on aspects of these interactions that are specifically relevant to nanostructures, where confinement of an electron system is important.

The Hamiltonian $H_I$ for a collection of nuclear spins in a solid divides naturally into five distinct terms:
\begin{equation}
 H_\mathrm{I} = H_\mathrm{Z}+H_\mathrm{hf}+H_\mathrm{orb}+H_\mathrm{dd}+H_\mathrm{Q}.
\end{equation}
Here, $H_\mathrm{Z}=-\sum_k \gamma_{j_k}I_k^z B$ describes the Zeeman energy in a magnetic field $B$ for a collection of nuclear spins of species $j_k$ at sites $k$ with associated gyromagnetic ratios $\gamma_{j_k}$ (we set $\hbar=1$, see Table \ref{tab:numerical} for numerical values of $\gamma_j$ for some relevant isotopes).  The hyperfine interaction between a collection of electron and nuclear spins is divided into two terms: $H_\mathrm{hf}=H_\mathrm{c}+H_\mathrm{a}$ where $H_\mathrm{c}$ is the isotropic (contact) part (see Sec. \ref{sec:contacthf}) and $H_\mathrm{a}$ gives the anisotropic hyperfine interaction (see Sec. \ref{sec:ahf}).  $H_\mathrm{orb}$ describes the coupling of nuclear spin to the electron orbital angular momentum (Sec. \ref{sec:orb}), $H_\mathrm{dd}$ gives the magnetic dipole-dipole coupling between a collection of nuclear spins (Sec. \ref{sec:dipdip}) and $H_\mathrm{Q}$ describes the quadrupolar interaction between nuclear spins and an electric-field gradient (Sec. \ref{sec:quadrupole}).

\subsection{Contact hyperfine interaction}\label{sec:contacthf}
The contact interaction was first derived by Fermi in 1930 \cite{Fermi1930a} to describe the spectroscopically observed hyperfine splitting of alkali metals.  The contact interaction is the most important term for describing electron-spin coherence in materials with a primarily s-type conduction band (see Sec. \ref{sec:decoherence}).  This includes all III-V semiconductors and silicon.
\begin{figure}[tb]
 \includegraphics[width=80mm]{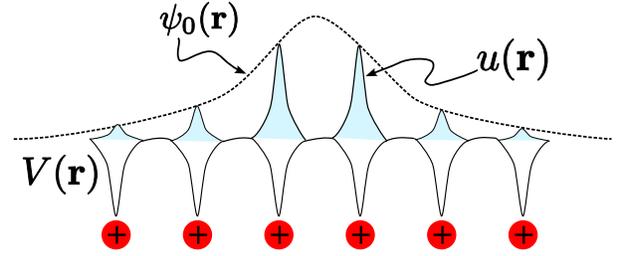}
\caption{\label{fig:envelope} Schematic diagram illustrating the electron envelope function $\psi_0(\mathbf{r})$, the $\mathbf{k}=0$ Bloch amplitude $u(\mathbf{r})$, and potential $V(\mathbf{r})$ created by positively charged nuclear cores.}
\end{figure}
For many electrons interacting with many nuclear spins in a solid, the contact interaction can be written generally as
\begin{equation}
 H_\mathrm{c} = -\frac{\mu_0}{4\pi}\cdot\frac{8\pi}{3}\gamma_S\sum_k \gamma_{j_k}\mathbf{S}(\mathbf{r}_k)\cdot\mathbf{I}_k,
\end{equation}
where $\gamma_S=-2\mu_B$ is the gyromagnetic ratio for a free electron\footnote{Due to the short-ranged nature of the contact interaction, the free-electron $g$-factor ($g\simeq2$) appears here, not the (renormalized) effective $g$-factor $g^*$.  See the discussion, e.g., by Yafet \cite{Yafet1961a}.}, $\mathbf{I}_k$ is the spin operator for a nucleus at atomic site $k$, and the electron spin density operator $\mathbf{S}(\mathbf{r})$ is given by
\begin{equation} \mathbf{S}(\mathbf{r})=\frac{1}{2}\sum_{s,s^\prime=\{\uparrow,\downarrow\}}\psi^\dagger_s(\mathbf{r})\pmb{\sigma}_{ss^\prime}\psi_{s^\prime}(\mathbf{r}), 
\end{equation}
with field operators defined by $\psi_\sigma(\mathbf{r})=\sum_n\phi_n(\mathbf{r})c_{n\sigma}$ and here
$c_{n\sigma}$ annihilates an electron in the state with spin $\sigma$ and single-particle orbital $\phi_n(\mathbf{r})$.  $\pmb{\sigma}_{ss'} = \bra{s}\pmb{\sigma}\ket{s'}$ gives the matrix elements for the vector of Pauli matrices.  The wave functions $\phi_n(\mathbf{r})$ are assumed to form a complete orthonormal set.

At low temperature, and neglecting possible valley degeneracy, a single electron confined to a quantum dot or bound to a donor impurity occupies a non-degenerate ground-state orbital $\bra{\mathbf{r}}\phi_0\rangle = \phi_0(\mathbf{r})$, which can be written (in the envelope-function approximation \cite{Winkler2003a}) as $\phi_0(\mathbf{r})=\sqrt{v_0}u(\mathbf{r}) \psi_0(\mathbf{r})$, where $v_0$ is the atomic volume\footnote{Since $v_0$ is chosen to be the atomic volume (rather than the primitive-cell volume), the Bloch amplitudes are normalized over a unit cell $\Omega$ according to: $\int_{\Omega} d^3 r |u(\mathbf{r})|^2=n_a$, where $n_a$ is the number of atoms in $\Omega$, consistent with refs. \cite{Coish2004a,Coish2008a,Fischer2008a}.  This normalization has the advantage that the resulting value of $A^j$ is independent of $n_a$.  However, it is different from that adopted by other authors \cite{Paget1977a,Merkulov2002a,Liu2007a,Eble2009a}, who take $\int_{\Omega} d^3 r |u(\mathbf{r})|^2=1$, resulting in a hyperfine coupling constant ${A^j}^\prime = A^j/n_a$.  In III-V semiconductors, the appropriate factor of $n_a=2$ for a Zincblende primitive cell should be taken into account when comparing $A^j$ values calculated using the two distinct normalizations.}, $u(\mathbf{r})$ is the lattice-periodic $\mathbf{k}=0$ Bloch amplitude and $\psi_0(\mathbf{r})$ is the slowly-varying ground-state envelope function (see Fig. \ref{fig:envelope}).  When the electron orbital level spacing is large compared to $k_B T$ and the scale of the hyperfine coupling, the electron-nuclear spin system will be well-described by the effective Hamiltonian, projected onto the ground-state orbital:
\begin{equation}
H_c^{\mathrm{eff}} = \left\langle \phi_0\right| H \left|\phi_0\right\rangle = \sum_k A_k\mathbf{S}\cdot\mathbf{I}_{k},
\label{eq:FermiHamiltonianDot}
\end{equation}
where $A_k=A^{j_k} v_0|\psi_0(\mathbf{r}_k)|^2$, and 
\begin{equation}
 A^{j_k} = -\frac{\mu_0}{4\pi}\cdot\frac{8\pi}{3}\gamma_S\gamma_{j_k}|u(\mathbf{r}_k)|^2
\end{equation}
is the total hyperfine coupling constant for a nucleus of species $j_k$ at position $\mathbf{r}_k$ within a crystal unit cell.  The free-electron gyromagnetic ratio is always negative ($\gamma_S<0$), but the nuclear gyromagnetic ratio $\gamma_j$ can take either sign (see Table \ref{tab:numerical}), leading to a hyperfine coupling constant that is either positive or negative \cite{Schliemann2003a}.

The hyperfine coupling constant $A^j$ depends on both the nuclear isotope $j$ (through $\gamma_j$) and electronic properies of the relevant material (through the Bloch amplitude $u(\mathbf{r}_j)$).  The dependence of $A^j$ on the electronic structure makes estimates of hyperfine coupling constants particularly challenging.  When direct experimental values for $A^j$ are unavailable, it is often necessary to rely on comparisons to related materials \cite{Paget1977a}, tight-binding methods \cite{Kohn1955a,Gueron1964a,Fischer2008a}, or \emph{ab initio} calculations \cite{Feller1983a,Overhof2004a,Yazyev2008a} for small clusters.
 
In a material containing several different nuclear isotopic species $j$, each with associated abundance $\nu_j$, it is common to define an average hyperfine coupling constant.  Here, we take the r.m.s. average:
\begin{equation}
 A = \sqrt{\sum_j \nu_j (A^j)^2}.
\end{equation}

Gated lateral quantum dots are typically formed in a GaAs two-dimensional electron gas (2DEG).  In GaAs, the three naturally occurring isotopes, $^{69}$Ga, $^{71}$Ga, and $^{75}$As, all have nuclear spin $I=3/2$ and the relative abundances are $\nu_{^{69}\mathrm{Ga}}  =  0.3$, $\nu_{^{71}\mathrm{Ga}}  =  0.2$, and $\nu_{^{75}\mathrm{As}}  =  0.5$.  Using these abundances with the coupling constants listed in Table \ref{tab:numerical} gives an r.m.s. coupling strength $A=85\,\mu$eV.  This coupling is rather strong; a fully-polarized nuclear spin system leads to an effective magnetic (Overhauser) field of $|IA/g^*\mu_B|\approx 5\,\mathrm{T}$ in GaAs (using the bulk value of $g^*=-0.4$).

\begin{table} 
\begin{center}
  \begin{tabular}{ |l || c || c || c || c |}
    \hline
     & I & $\gamma_j\;(\mathrm{rad}\,\mathrm{T}^{-1}\,s^{-1}$) & $A^j$ ($\mu$eV)  & $Q_j$ (mb)\\ \hline
    $^{69}$Ga & 3/2 & $6.43\times 10^{7}$ &  74 \cite{Paget1977a}$^\dagger$   & 171 \cite{Pyykko2008a}\\ \hline
    $^{71}$Ga & 3/2 & $8.18\times 10^{7}$ &  96 \cite{Paget1977a}$^\dagger$   &  107 \cite{Pyykko2008a}\\ \hline
    $^{75}$As & 3/2 & $4.60\times 10^{7}$ &  86 \cite{Paget1977a}$^\dagger$ &  314 \cite{Pyykko2008a}\\ \hline
    $^{113}$In & 9/2 & $5.88\times 10^{7}$ &  110 \cite{Liu2007a}$^\dagger$ &   759 \cite{Pyykko2008a}\\ \hline
    $^{115}$In &9/2 & $5.90\times 10^{7}$ &  110 \cite{Liu2007a}$^\dagger$ &   770 \cite{Pyykko2008a}\\ 
    \hline
    $^{13}$C & 1/2 & $6.73\times 10^{7}$ & - & 0\\
    \hline
    $^{29}$Si & 1/2 & $-5.32\times 10^{7}$ & - & 0\\
\hline
    $^{14}$N & 1 & $1.93\times 10^{7}$ & - & 20.44 \cite{Pyykko2008a}\\
\hline
   $^{15}$N & 1/2 & $-2.71\times 10^{7}$ & - & 0\\
\hline
  \end{tabular}
\end{center}
$^\dagger$ See footnote 2, below.
\caption{\label{tab:numerical}
Nuclear spin, gyromagnetic ratios, contact hyperfine coupling strengths in $\mathrm{In}_x\mathrm{Ga}_{1-x}\mathrm{As}$, and quadrupole moments for some isotopes that appear in quantum dots and nitrogen vacancy centers in diamond.  Note that 1 mb = $10^{-31}$ $\mathrm{m^2}$.}
\end{table}

\subsection{Anisotropic hyperfine}\label{sec:ahf} While the contact interaction is dominant in the s-type conduction band of III-V semiconductors and silicon, bands primarily composed of p-orbitals (e.g., the valence band in III-V semiconductors \cite{Fischer2008a}, or the $\pi$-orbitals in carbon nanotubes and graphene \cite{Fischer2009a}) have a wave function that vanishes at the nuclear sites, resulting in a vanishing contact interaction.   In this case, the largest sources of electron-nuclear coupling are provided by the anisotropic hyperfine interaction (see below) and the coupling to orbital angular momentum (Sec. \ref{sec:orb}).  The anisotropic interaction is also important for defects in diamond and molecular magnets, where the electronic wavefunctions have low symmetry.

The anisotropic hyperfine interaction for a collection of electron and nuclear spins can be written most generally in terms of a Hamiltonian density: $H_a =\int d^3 r H_a(\mathbf{r})$, where
\begin{equation}\label{eq:Ha}
 H_a(\mathbf{r}) = \sum_{k}\mathbf{S}(\mathbf{r})\cdot\stackrel{\leftrightarrow}{\mathbf{T}}_k(\mathbf{r})\cdot\mathbf{I}_k,
\end{equation}
and $\stackrel{\leftrightarrow}{\mathbf{T}}_k(\mathbf{r})$ is a traceless tensor with components given by
\begin{equation}
T_k^{\alpha\beta}(\mathbf{r})= \frac{\mu_0}{4\pi}\gamma_S\gamma_{j_k}\frac{\left(\delta_{\alpha\beta}-3\hat{n}_k^\alpha \hat{n}_k^\beta\right)}{|\mathbf{r}-\mathbf{r}_k|^3}.
\end{equation}
Here, $\alpha,\beta = \{x,y,z\}$, and $\mathbf{\hat{n}}_k=(\mathbf{r}-\mathbf{r}_k)/|\mathbf{r}-\mathbf{r}_k|$ is a unit vector, written in terms of the electron position operator $\mathbf{r}$.

As in Sec. \ref{sec:contacthf}, if the many-electron wave function is known, and the energy gap to the first excited state is large compared to the hyperfine coupling strength, we can form an effective Hamiltonian using Eq. (\ref{eq:Ha}) from the expectation value of the vector $\mathbf{S}(\mathbf{r})\cdot\stackrel{\leftrightarrow}{\mathbf{T}}_k(\mathbf{r})$ with respect to the electron state.  For a single electron in a localized orbital that is far from the nuclear sites, Eq. (\ref{eq:Ha}) reduces to the classical dipole-dipole coupling between the magnetic moments of the electron and nuclei.  However, typically the largest contribution comes from an `on-site' component, describing the electron density localized near the nucleus due to the Bloch amplitude $u(\mathbf{r})$ (see Fig. \ref{fig:envelope}) \cite{Yafet1961a}. A spherically symmetric distribution of electron density around the nucleus results in an average of Eq. (\ref{eq:Ha}) to zero.  Thus, for an electron in an s-type conduction band, the on-site component of (\ref{eq:Ha}) vanishes and the remaining contributions will be much weaker long-ranged dipole-dipole interactions between the nuclear spin and electron spin density at distant atomic sites.

\subsection{Nuclear-orbital interaction}\label{sec:orb}  The Pauli equation for a non-relativistic electron with momentum $\mathbf{p}$ in the presence of a vector potential $\mathbf{A}$ contains terms proportional to $\mathbf{A}\cdot\mathbf{p}$.  If $\mathbf{A}$ is generated by the magnetic moments of nuclear spins $\mathbf{I}_k$ located throughout a crystal, these terms can be rewritten as
\begin{equation}\label{eq:Horb}
 H_\mathrm{orb}=-\frac{\mu_0}{4\pi}\sum_{k}\gamma_S\gamma_{j_k} \frac{\mathbf{L}_{k}\cdot\mathbf{I}_k}{|\mathbf{r}-\mathbf{r}_k|^3}.
\end{equation}
Here, $\mathbf{L}_k$ is the operator for the total electron orbital angular momentum about the nuclear site $\mathbf{r}_k$.  Eq. (\ref{eq:Horb}) is particularly important for describing the electron-nuclear interaction for electrons in bands primarily composed of atomic orbitals with nonzero angular momentum.  For example, this term, along with Eq. (\ref{eq:Ha}), provides the dominant source of electron-nuclear interaction for electrons in the p-type valence band of III-V semiconductors  (i.e., for holes) \cite{Yafet1961a,Grncharova1977a,Fischer2008a,Eble2009a}.

\subsection{Nuclear dipolar interaction}\label{sec:dipdip}
In addition to the electron-nuclear interactions discussed above, the dipole-dipole interaction between individual nuclear spin magnetic moments plays an important role.  The dipole-dipole Hamiltonian can be written as:
\begin{equation}\label{eq:Hdd}
H_\mathrm{dd} = \sum_{k\ne l} \mathbf{I}_k\cdot\stackrel{\leftrightarrow}{\mathbf{T}}_{kl}\cdot\mathbf{I}_l,
\end{equation}
where the components of the tensor $\stackrel{\leftrightarrow}{\mathbf{T}}_{kl}$ are
\begin{equation}
 T_{kl}^{\alpha\beta} = \frac{\mu_0}{4\pi}\gamma_{j_k}\gamma_{j_l}\frac{\delta_{\alpha\beta}-3\hat{r}_{kl}^\alpha \hat{r}_{kl}^\beta}{2r_{kl}^3}.
\end{equation}
Here, $\mathbf{r}_{kl}=\mathbf{r}_k-\mathbf{r}_l$ and $\hat{\mathbf{r}}_{kl}=\mathbf{r}_{kl}/r_{kl}$.  Eq. (\ref{eq:Hdd}) contains terms that change the total $z$-component of spin, and can therefore lead to local spin-flips.  However, in a moderate magnetic field (larger than a few Gauss), only the secular part of Eq. (\ref{eq:Hdd}) (that which commutes with the nuclear Zeeman term $H_Z$) contributes:
\begin{equation}\label{eq:HddSecular}
 H_\mathrm{dd}^\mathrm{sec.}= \sum_{k\ne l} d_{kl}I_k^z I_l^z-\frac{1}{2}\sum_{k\ne l,(j_k = j_l)}d_{kl}I_k^+I_l^-,
\end{equation}
where $d_{kl} = \left(\mu_0/4\pi\right)\gamma_{j_k}\gamma_{j_l}\left(1-3\cos^2\theta_{kl}\right)/2r_{kl}^3$. The second sum in Eq. (\ref{eq:HddSecular}) is restricted to run over pairs of sites with the same isotopic species and $\theta_{kl}$ is the angle between the magnetic field and the vector $\mathbf{r}_{kl}$.   While Eq. (\ref{eq:HddSecular}) conserves the total $z$-component of nuclear spin, the second term gives rise to flip-flops between nuclear spins of the same species at different sites. In combination with the hyperfine interaction, these flip-flops can cause electron-spin decoherence through spectral diffusion  \cite{Klauder1962a,DeSousa2003a} (see Sec. \ref{sec:decoherence}), and can redistribute nuclear spin polarization through nuclear spin diffusion (see Sec. \ref{sec:DNP}).

\subsection{Nuclear quadrupolar interaction}\label{sec:quadrupole}
The intrinsic electric dipole moment of a nucleus, if nonzero, must be extremely small \cite{Purcell1950a,Commins2007a}.  Nuclear spins are therefore immune to interaction with constant electric fields.  However, a nucleus with spin $I>1/2$ does have a finite electric quadrupole moment, and can therefore couple to electric field \emph{gradients} through the electric quadrupole term due to a nonuniform electrostatic potential $V(\mathbf{r})$ \cite{Abragam1962a}: 
\begin{equation}\label{eq:HQ}
 H_{Q} = \sum_k\sum_{\alpha\beta}V_{\alpha\beta}^k Q^{\alpha\beta}_k.
\end{equation}
We use the notation $V_{\alpha\beta}^k=\left.\left<\frac{\partial^2 V(\mathbf{r})}{\partial x^\alpha\partial x^\beta}\right>\right|_{\mathbf{r}=\mathbf{r}_k}$, where $\left<\cdots\right>$ indicates an expectation value with respect to the electron system and the quadrupole tensor is given by
\begin{equation}
 Q^{\alpha\beta}_k = eQ_{j_k}\frac{\left[\frac{3}{2}(I^\alpha_k I^\beta_k+I^\beta_k I^\alpha_k)-\delta_{\alpha\beta}I^{j_k}(I^{j_k}+1)\right]}{6 I^{j_k}(2I^{j_k}-1)}.
\end{equation}
Values of the quadrupole moment $Q_j$ for several important isotopes $j$ are given in Table \ref{tab:numerical}.  In a crystal with cubic symmetry, $V_{xx}=V_{yy}=V_{zz}$, the electric field gradient (and hence the quadrupolar term) must vanish due to Laplace's equation ($V_{xx}+V_{yy}+V_{zz}=0$) \cite{Abragam1962a}.  Crystal strain due to a semiconductor heterostructure, dopants, or defects will, however, give rise to nonzero electric-field gradients at the positions of the nuclei, giving significant values for the quadrupolar splitting.  A strong quadrupolar splitting has been seen in nanostructures, resulting in allowed multiple-quantum transitions with $\Delta m=\pm 2$ \cite{Salis2001a,Salis2001b}, and a measured shift in the nuclear spin resonance line of $\gtrsim 10\,\mathrm{kHz}$ in a GaAs 2DEG \cite{Yusa2005a}.

Non-secular terms in Eq. (\ref{eq:HQ}) can lead to an important spin-lattice relaxation mechanism (nuclear spin flips).  However, in a small applied magnetic field, the remaining (secular) part of $H_Q$ preserves the component of nuclear spin along the magnetic field.  Assuming axial symmetry for the potential about some direction $\hat{n}$, the secular quadrupolar term is \cite{Abragam1962a}:
\begin{equation}\label{eq:HQSec}
 H_{Q}^\mathrm{sec.} = \frac{1}{4}\sum_k\nu_Q^k f(\theta)\left((I_k^z)^2-\frac{1}{3}I^{j_k}(I^{j_k}+1)\right),
\end{equation}
where $f(\theta)=\left(3\cos^2\theta-1\right)$ and $\theta$ is the angle between $\hat{n}$ and the applied magnetic field (along $z$).  The quadrupolar coupling strength is
\begin{equation}\label{eq:nuQ}
 \nu_{Q}^k = \frac{3eV_{nn}^k Q_{j_k}}{8I^{j_k}(2I^{j_k}-1)}.
\end{equation}
Here, $V_{nn}^k=\left.\left<\hat{n}\cdot\pmb{\nabla}(\hat{n}\cdot\pmb{\nabla}V(\mathbf{r}))\right>\right|_{\mathbf{r}=\mathbf{r}_k}$ is the (negative) electric field gradient along $\hat{n}$.

For a single electron in a spherically symmetric $s$-orbital, the electric field gradient due to the electron charge distribution vanishes at the site of the nucleus.  For states of finite angular momentum ($p$-, $d$-, etc.), there is a nonvanishing contribution, in general.  The order of magnitude of this interaction is, however, typically small compared to the interactions given in Eqs. (\ref{eq:FermiHamiltonianDot}), (\ref{eq:Ha}), and (\ref{eq:Horb}).   To estimate the size of $\nu_Q^k$, we again employ the envelope function approximation $\phi(\mathbf{r})=\sqrt{v_0}u(\mathbf{r})\psi(\mathbf{r})$, which gives
\begin{equation}
  \nu_Q^k=E_Q^{j_k}v_0|\psi(\mathbf{r}_k)|^2.
\end{equation}
Here, $E_Q^{j_k}$ is given by Eq. (\ref{eq:nuQ}), but with the expectation value in $V_{nn}$ taken with respect to the Bloch amplitude $u(\mathbf{r})$ over a single unit cell.  To see the typical size of this term, we estimate the quadrupolar splitting for a $^{69}$Ga nuclear spin interacting with a heavy hole in the valence band of GaAs (due purely to the electric field gradient due to the electron density: $V(\mathbf{r})=e/4\pi\epsilon_0r$) as
\begin{equation}\label{eq:EQ}
 E_Q^{^{69}\mathrm{Ga}} = \frac{e^2}{4\pi\epsilon_0}\frac{Q_{^{69}\mathrm{Ga}}}{8}\left<\frac{3\cos^2\theta-1}{r^3}\right>_{4p}\simeq -0.01\,\mu eV.
\end{equation}
This value can be compared directly with the strength of the combined anisotropic hyperfine and orbital contributions for a hole in GaAs, giving a coupling strength on the order of \cite{Fischer2008a} $A_{h}\sim 10\,\mu eV$.  In Eq. (\ref{eq:EQ}), we have used that $\left<1/r^3\right>_{4p}=1/192(a_B^\mathrm{eff})^3$ and $\left<\cos^2\theta\right>_{4p}=1/5$ for a hydrogenic $4p$ orbital, with an effective Bohr radius ($a_B^\mathrm{eff} = 8.5\times 10^{-12}\,\mathrm{m}$ for Ga \cite{Clementi1963a}) that accounts for screening due to the core-shell electrons.  We emphasize that Eq. (\ref{eq:EQ}) estimates only the on-site electronic contribution to the quadrupolar splitting and that the overall splitting due to lattice strain can be significant.

The primary effect of the secular quadrupole term (Eq. (\ref{eq:HQSec})) is to give an unequal spacing to the nuclear Zeeman levels in an applied magnetic field.  As a consequence, it is possible to individually address transitions between, e.g., $m=1/2\leftrightarrow-1/2$ and $m=1/2\leftrightarrow 3/2$ states with different excitation frequencies, allowing for full control of the single-spin Hilbert space and the execution (in principle) of quantum algorithms \cite{Leuenberger2002a}.  An inhomogeneous quadrupolar splitting can also suppress dipolar nuclear spin flip-flops due to the secular dipole-dipole coupling (Eq. (\ref{eq:HddSecular})) when $|\nu_Q^k-\nu_Q^l|\gtrsim |d_{kl}|$.  This effect can significantly reduce nuclear spin diffusion in a strained sample \cite{Maletinsky2009a} (see Sec. \ref{sec:DNPLimits}, below).

\section{Dynamic nuclear polarization}\label{sec:DNP}

Hyperfine coupling to electron spins can serve as a pathway for the nuclear spin system to relax to its thermal equilibrium state, or for the production of highly non-equilibrium dynamic nuclear polarization (DNP) states when certain external forcing mechanisms are applied. DNP was first observed by Carver and Slichter in 1953 \cite{Carver1953a}, who confirmed the theory of Overhauser \cite{Overhauser1953a} for microwave-driven polarization of nuclei in metals. Later seminal work on DNP was carried out by Abragam and Proctor \cite{Abragam1958a} on the so-called 'solid-effect' involving electronic defect centers in dielectric materials. The situation in semiconductor quantum dots more closely resembles that of conduction electrons in metals, since a single electron is simultaneously coupled to $N \sim 10^4 - 10^6$ nuclear spins through the contact hyperfine interaction (Eq. \ref{eq:FermiHamiltonianDot}). 

Although here we focus on DNP, we note that an intriguing alternative to \emph{dynamic} polarization is the possibility of a nuclear-spin ferromagnetic phase transition below the Curie temperature $T_c$ due to a coupling mediated by the hyperfine interaction with an electron system, first predicted for metals by Fr\"ohlich in 1940 \cite{Frohlich1940a}.  Recent theory suggests that $T_c$ for this transition may reach reasonable dilution-refrigerator temperatures in strongly correlated low-dimensional systems \cite{Simon2007a,Simon2008a,Braunecker2009a}, but this transition has yet to be verified experimentally.

\subsection{Optical pumping of nuclear spins in quantum dots}

\begin{figure}[tb]
\includegraphics[width=80mm]{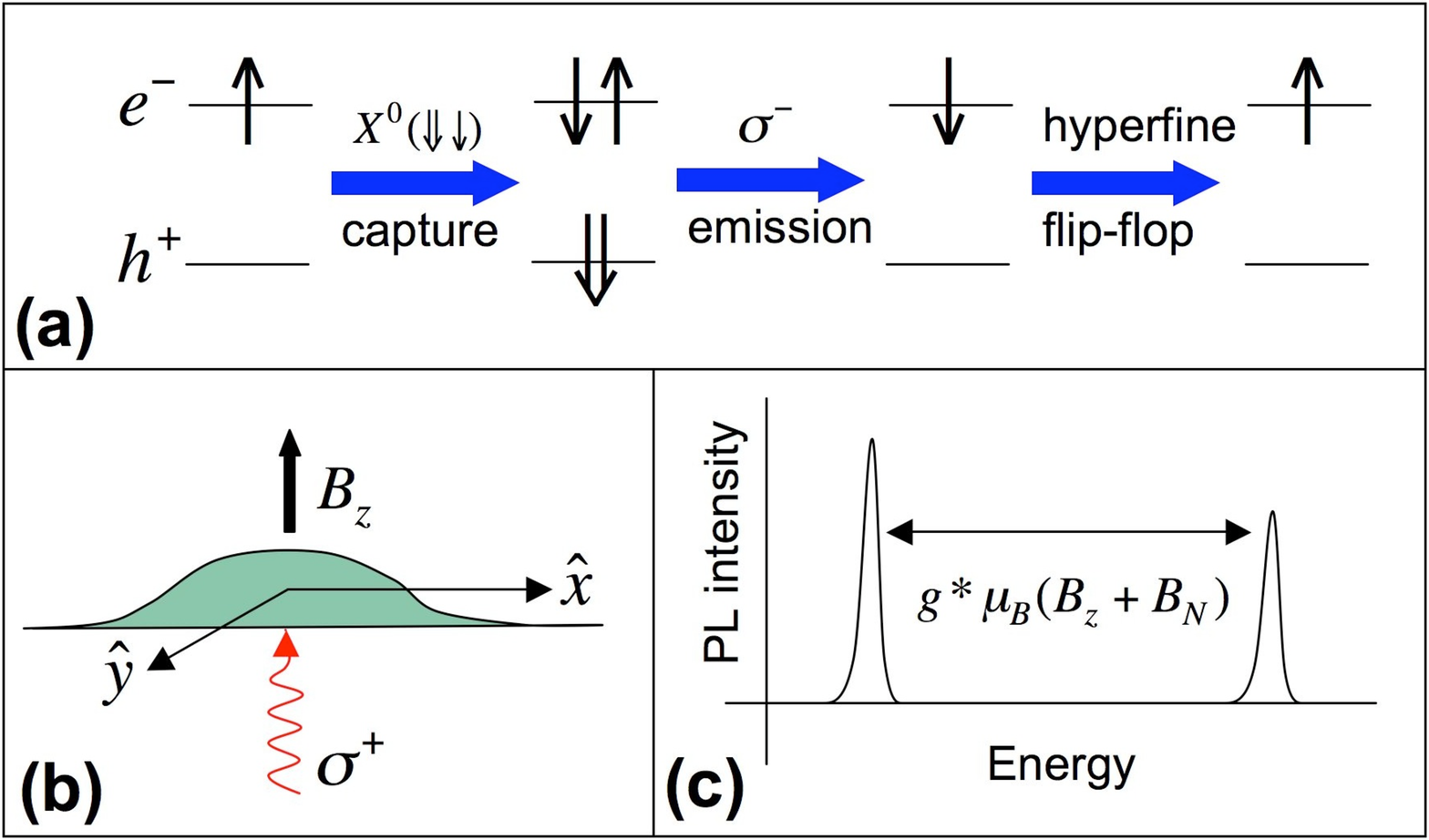}
\caption{\label{fig:optpumping} (a) Example of a process leading to electron and nuclear spin pumping. A singly occupied dot ($\uparrow$) captures a dark exciton ($X^{0}$), followed by recombination leaving a spin down electron; the optically pumped electron can exchange angular momentum with a nuclear spin mediated by the hyperfine flip-flop process. (b) Longitudinal applied field geometry for observing the Overhauser shift in a self-assembled quantum dot, and (c) schematic of Zeeman-split photoluminescence peaks, as observed in the experiments of Bracker et al. \cite{Bracker2005a}.}
\end{figure}

DNP was first observed in single quantum dots via optical pumping of electron spins \cite{Brown1996a}. Optical pumping can be thought of as a two-step process: first, excitation by circularly polarized light transfers angular momentum to electron spins, creating a net electronic polarization; second, angular momentum is transferred to the nuclear spin system via the hyperfine interaction together with processes that either remove or relax the electron spin. The nuclear spin polarization then acts back on the electron spin through an effective magnetic field, the Overhauser field \cite{Overhauser1953a}:
\begin{equation}
B_N=\frac{\sum_k A_k \left<I_{k}^z\right>}{g^*\mu_B},
\end{equation}
where here, $\left<\cdots\right>$ indicates an expectation value with respect to the nuclear spin state.  $B_N$ has the effect of either increasing or decreasing the electronic Zeeman spin splitting, depending on its sign, and so can be observed spectroscopically. In Bracker et al. \cite{Bracker2005a}, Zeeman splittings were observed in photoluminescence spectra of excitons in a single charge-tunable self-assemble quantum dot in a longitudinal magnetic field (see Fig. \ref{fig:optpumping}). When pumped with circularly polarized light, Bracker et al. report Overhauser shifts of the splittings as large as 81 $\mu eV$, corresponding to a nuclear polarization $P_N=81\,\mu eV/IA = 60\%$ ($IA=135\,\mu eV$ for GaAs has been estimated in Ref. \cite{Paget1977a}). By controlling the charge state of the dot prior to excitation, they are able to measure the electronic and nuclear polarizations for neutral ($X^0$) and charged ($X^+, X^-$) excitons, demonstrating that the nuclear polarization tracks the electron polarization in each case, and that both can be tuned with applied bias. In these and earlier experiments, it was assumed that an external magnetic field larger than the nuclear dipole-dipole couplings was necessary for DNP, so that non-spin-preserving (i.e. nonsecular) terms  in the dipolar Hamiltonian would be suppressed \cite{Gammon2001a}. However, Lai et al. \cite{Lai2006a} demonstrated that DNP could be achieved by optical pumping in the \emph{absence} of an external magnetic field, due to the effective magnetic field of the polarized electrons (Knight field) acting on the nuclear spins, suppressing the nonsecular dipolar interactions and providing a quantization axis along which the nuclei can polarize. Lai et al. estimate that this Knight field to be $\sim 100-200$ Gauss for a fully polarized electron, about an order of magnitude larger than the characteristic local dipolar field. It was left as an open question why the maximal nuclear polarization observed in this regime is only $\sim 10-15\%$ \cite{Lai2006a}. Maletinsky et al. \cite{Maletinsky2007a} studied the buildup and decay of DNP in this zero-field and low-field regime, and found that a resident electron in the dot could relax the nuclear polarization on the millisecond timescale. This was attributed to two possible mechanisms: the indirect coupling of nuclear spins via the electron (combined with the effect of the nonsecular dipolar terms at very low fields), and depolarization of the electron due to cotunneling processes which exchange the resident electron with one in the reservoir. Here, the cotunneling timescale is estimated to be $\sim 20$ nanoseconds  \cite{Maletinsky2007a} (a later work confirmed the cotunneling mechanism by investigating samples with various barriers between the dot and reservoir \cite{Latta2009a}). By removing the electron with a gate pulse, or going to larger magnetic fields, Maletinsky et al. showed much prolonged nuclear decay times up to seconds or minutes. In a second paper, Maletinsky et al. \cite{Maletinsky2007b} study the dependence of the optically excited DNP on external magnetic field $B_{ext}$ from $-2$ to $+2$ Tesla. They found a magnetic hysteresis in the Overhauser shift indicative of a bistability, and derived a semiclassical rate equation model to explain this based on the dependence of the electron-mediated nuclear relaxation rate on the total electronic Zeeman splitting (i.e. the sum of external field and Overhauser field). This dependence of the nuclear pumping rate on the Overhauser field leads to non-linear dynamics of the combined electron-nuclear spin system. The maximal DNP pumping rate occurs when the total electronic Zeeman energy is zero, i.e. when $B_{N}=-B_{ext}$; a further increase of $B_{ext}$ leads to a drop in $|B_N|$. The model of Maletinsky et al. predicts that the maximal DNP is limited by the ratio of the nuclear polarization decay rate (e.g. due to spin diffusion out of the dot) to the timescale for the nuclear and electron spin systems to reach thermal equilibrium (i.e. the electron-mediated nuclear relaxation rate).  The latter timescale is proportional to $N^2/(fA^2)$, where $N$ is the number of nuclear spins in the dot and $f$ is the fraction of time the dot is occupied by an electron. 

Braun et al. \cite{Braun2006a} observed a similar magnetic field dependence, additionally saw bistable behavior as a function of the electron spin polarization, and explained both with a semiclassical model similar to that of Maletinsky et al. Regarding maximal polarization, Braun et al. emphasize the likely competition between too large an external field making electron-nuclear spin flips too costly for efficient pumping, and too low an external field in which nuclear decay processes such as quadrupolar relaxation are not efficiently suppressed \cite{Deng2005a}. In their experiments on InGaAs dots, optimal pumping of DNP is found to occur at fields between $1.5$ and $2.5$ Tesla. Urbaszek et al. \cite{Urbaszek2007a} performed similar experiments on InGaAs dots with a single positively charged exciton ($X^+$) as a function of temperature from $2$K to $55$K, finding a surprising increase in nuclear polarization as temperature increases. This is attributed to a broadening of the electronic Zeeman levels increasing the rate of electron-nuclear spin flip-flops. Recent work by Latta et al. has demonstrated bi-directional polarization controlled by setting laser detuning on either side of the dot ($X^{-}$) resonance \cite{Latta2009a}. The nuclear spins polarize so as to maintain the resonance condition, thereby ``dragging" the resonance. Such a feedback mechanism is expected to narrow the nuclear spin distribution (suppress fluctuations) as long as the feedback response is faster than the random nuclear fluctuations \cite{Latta2009a}.

\subsection{Electrically controlled DNP in double quantum dots}

\begin{figure}[tb]
\includegraphics[width=80mm]{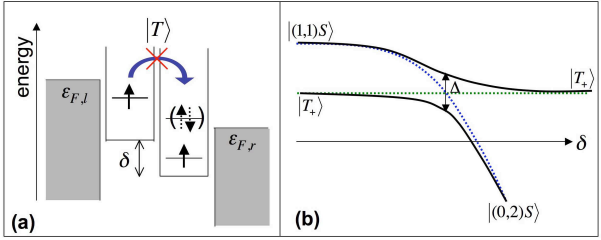}
\caption{\label{fig:PSB} Pauli spin blockade and nuclear pumping. (a) Band diagram of the double-dot spin blockade setup, showing that transport is blocked for triplet states $\ket{T}$, but can proceed for the singlet $\ket{S}$. $\delta$ denotes the energy detuning of the dots, and $\epsilon_F$ the Fermi energy in the leads. (b) Schematic energy diagram versus detuning, showing an anti-crossing between triplet $\ket{T_+}=\ket{\uparrow\uparrow}$ and singlet states, with a splitting $\Delta$ arising from the hyperfine interaction. Such a situation was exploited in references \cite{Foletti2008a} and \cite{Petta2008a} to generate DNP one electron-nuclear flip-flop at a time by adiabatic passage from initial state $\ket{(0,2)S}$ to the anti-crossing to allow hyperfine mixing.}
\end{figure}

\begin{figure}[tb]
\includegraphics[width=80mm]{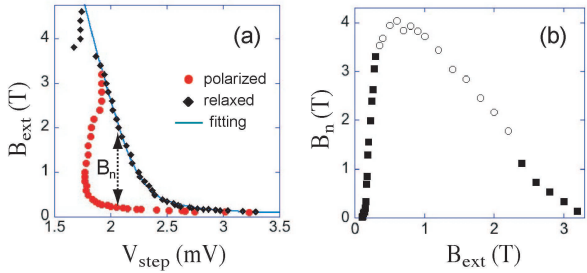}
\caption{\label{fig:DNP} Figure from reference \cite{Baugh2007a} showing (a) how the Overhauser field $B_{n}$ is extracted from the positions of current steps in the dc magnetotransport data, and (b) the Overhauser field as a function of external magnetic field $B_{ext}$.}
\end{figure}

The seminal observation of the two-electron Pauli spin blockade in a vertically-coupled double quantum dot by Ono \emph{et al.} \cite{Ono2002a,Ono2004a} laid the foundation for much subsequent work using transport measurements to study electron and nuclear spin dynamics in quantum dots. Consider two electrons in adjacent dots: if the potential of the left dot is raised until it is larger than the charging energy required add a second electron to the right dot, the left electron will tunnel onto the right dot to minimize total energy. However, this process is prohibited due to the Pauli exclusion principle if the two electrons form a spin triplet state; the same orbital in the right dot cannot be doubly occupied unless the electrons form a spin singlet (see Fig. \ref{fig:PSB}a). Magneto-transport measurements carried out in the spin blockade regime of an InGaAs vertical double dot device revealed current features exhibiting magnetic hysteresis, instabilities and low frequency (e.g. $\lesssim$ 1 Hz) oscillations \cite{Ono2004a}. This behavior was attributed to DNP, but the exact mechanism was not well understood, particularly since electrons in the leads are expected to be completely unpolarized. Similar hysteresis and bistabilities (though not coherent oscillations) were later observed independently in GaAs lateral quantum dots \cite{Koppens2005a}.  Subsequent work by Baugh et al. \cite{Baugh2007a} quantified the degree of polarization in vertical GaAs double-dot devices as a function of external magnetic field and proposed a mechanism to explain the behavior. Baugh et al. reported a maximal Overhauser field of $\sim 4$ Tesla, corresponding to a polarization $\sim 40\%$ (Fig. \ref{fig:DNP}). Here, DNP occurs when one of the blockaded spin triplets ($\ket{T_-}=\ket{\downarrow\downarrow}$) comes close to degeneracy with the spin singlet branch that has mostly $\ket{S(1,1)}$ character, where $(n, m)$ represents the number of electrons in the (left, right) dot. When the energy difference between $\ket{T_-}$ and $\ket{S(1,1)}$ becomes small, the hyperfine interaction drives the transition $\ket{T_-}\rightarrow\ket{S(1,1)}$, accompanied by a nuclear spin flip to conserve angular momentum. The state $\ket{S(1,1)}$ rapidly relaxes to the lower energy state $\ket{S(0,2)}$, and finally to the charge state $(0,1)$ as an electron tunnels out into the right lead. Since the leads are unpolarized, the probabilities are equal for the system to be blockaded in any of the triplet states, so that nuclear polarization can only accumulate if the other triplet states $\ket{T_0}$ and $\ket{T_+}$ have suitably short lifetimes due to processes \emph{unrelated} to the hyperfine interaction. In these experiments, strong cotunneling due to relatively transparent dot-lead tunnel barriers serves this function \cite{Qassemi2008a}. The $m_s=\pm1$ triplet levels are shifted by the average Overhauser field of the two dots, and in the experiments of Baugh et al., this leads to a shift in detuning of the position of a current step observed in dc transport. By plotting the step position as a function of external field for both polarized and unpolarized states, the Overhauser field can be extracted as in Fig. \ref{fig:DNP}.

Electrical control of DNP was taken a step further in the work of Petta et al. \cite{Petta2008a} and Foletti et al. \cite{Foletti2008a} in GaAs lateral quantum dots. They utilized the singlet-triplet anticrossing shown in Fig. \ref{fig:PSB}b to generate DNP by applying a voltage cycle to load electrons into the $\ket{(0,2)S}$ state and then bring them adiabatically to the $S/T_+$ anticrossing to induce an electron-nuclear flip-flop. In this way one nuclear spin is flipped per cycle, and the Overhauser shift monitored by the position of the $S/T_+$ anticrossing with respect to detuning. Petta et al. showed that one version of this cycle allows the steady-state polarization to be set by choosing the detuning at which the adiabatic return passage ends; when the $S/T_+$ anticrossing coincides with this detuning, buildup of polarization stops \cite{Petta2008a}. Foletti et al. studied a similar sequence wherein the reload step is removed so that the same pair of electrons is retained throughout \cite{Foletti2008a}. They observed some oscillation of the nuclear polarization as a function of external field, and attributed this to an interplay between the cycle time and the Larmor frequencies of the nuclear spins. In both cases, the maximum polarization reached was of order $\sim 1-2 \%$. 

Recent experiments using electron spin resonance in the spin blockade regime have demonstrated resonance dragging due to DNP \cite{Vink2009a} similar to the recent observation in optically pumped dots by Latta et al. Finally, several important theoretical works have recently been devoted to the effects of DNP on leakage current \cite{Inarrea2007b} and hysteresis \cite{Inarrea2007a} in the spin blockade regime, DNP in the presence of spin relaxation \cite{Rudner2007a}, resonant electric- \cite{Rudner2007b} and magnetic-field excitation \cite{Danon2008a}, and the creation of dynamical stabilities under pumping \cite{Danon2009a}.

\subsection{Limits to polarization}\label{sec:DNPLimits}
The nuclear polarizations that can be achieved by these methods are typically limited either by loss rates (e.g. intrinsic nuclear spin-lattice relaxation or spin diffusion \cite{Ramanathan2008a} out of the dot) or by a suppression of the hyperfine flip-flop process as polarization is built up. For example, the cyclical adiabatic methods used in the double dot system \cite{Petta2008a,Foletti2008a} could in principle give a polarization rate independent of the polarization state, so that maximal polarization is only determined by the loss rate. The loss rate observed by Petta et al. could be explained by spin diffusion perpendicular to the 2DEG plane, and was much faster than the polarization rates of the employed cycles, limiting polarization to $\sim 1\%$. If spin diffusion were suppressed or the polarization cycle time greatly reduced, this method could yield polarizations near unity. In optically pumped self-assembled dots, spin diffusion can be eliminated for isotopes that occur only in the dot material and not in the surrounding matrix, from a non-uniform Knight field due to site-dependent hyperfine coupling constants $A_k\propto|\psi(\mathbf{r}_k)|$, or due to a non-uniform quadrupolar splitting $\nu_Q^k$, yielding exceedingly long polarization storage times \cite{Greilich2007a,Maletinsky2009a}. The challenge there is to optically produce $100\%$ electronic polarization, and to suppress relaxation due to non-secular nuclear terms at high field while keeping the electron-nuclear flip-flop rates sufficiently large.  Another promising method to extend the lifetime of a polarized nuclear spin system is to perform a sequence of rapid measurements on the nuclear Overhauser field (the quantum Zeno effect) \cite{Klauser2008a}.  Experiments have yet to demonstrate a robust Zeno effect in practise.

\section{Electron spin decoherence}\label{sec:decoherence}

Historically, electron-spin decoherence has typically been evaluated within Bloch-Redfield theory \cite{Bloch1957a,Redfield1957a}.  Bloch-Redfield theory is valid in the limit where an electron interacts weakly with an environment (validating a weak-coupling expansion), which itself has a short correlation time (allowing a Markov approximation).  The result of Bloch-Redfield theory is particularly simple; the components of electron spin along and transverse to an applied magnetic field decay exponentially with the time scales $T_1$ and $T_2$, respectively.  While the $T_1$ time for localized spins in a large magnetic field is typically limited by spin-orbit interaction and phonon emission (a mechanism for which Bloch-Redfiled theory applies) \cite{Pines1957a,Khaetskii2001a,Golovach2004a,Elzerman2004a,Kroutvar2004a}, the transverse-spin decay time is often limited by electron-nuclear interactions \cite{Klauder1962a,Khaetskii2002a,Merkulov2002a,DeSousa2003a,Coish2004a,Abe2004a,Petta2005b,Koppens2008a}.  Due to the significant strength of the hyperfine interaction (see Sec. \ref{sec:contacthf}), a weak-coupling expansion is typically not possible, and because of the relatively long nuclear correlation time $\tau_c$, a Markov approximation is also typically invalid, leading generically to non-exponential (non-Markovian) decay of spin correlations \cite{Khaetskii2002a,Coish2004a,Coish2005a}.

Determining the quantum dynamics of a `central' electron spin interacting with an environment of other `bath' nuclear spins is a complicated many-body problem, which has historically led authors to seek phenomenological solutions \cite{Anderson1953a,Klauder1962a}.  This previous work gives important insight into the major mechanisms of the decay processes.  However, phenomenological theories may not be sufficiently accurate to understand decoherence at the level required for fault-tolerant quantum information processing \cite{Preskill1998a,Steane2003a,Knill2005a}.  Moreover, previous theory has focused on the experimental system that was relevant at the time; an ensemble of decohering spins, with associated inhomogeneity.  New experiments now allow for the controlled creation and measurement of single-spin coherence \cite{Jelezko2004a,Petta2005b,Koppens2008a}, opening the door for new methods of coherence preservation that were not available until very recently.

The traditional view of spin decoherence emphasizes that spin ensembles suffer from inhomogeneous broadening (due, e.g., to a random local magnetic field), resulting in a rapid free-induction decay (decay in the absence of spin echo pulses).  Ideally, spin echoes remove the effects of inhomogeneities in an ensemble, giving the `true' decay time for a single spin.  Although it is certainly true that inhomogeneities in spin ensembles can result in rapid decay, it is also possible for spin echoes to refocus decoherence of a single spin interacting with a quantum-mechanical environment, extending the decay time for a single spin.  This fact makes it necessary to consider both problems (free-induction decay and decay under spin echoes) independently, even in the case of single-spin decoherence.

In the context of quantum information processing, a finite spin-rotation (qubit gating) time $t_g$ typically results in an error per gate $\propto t_g/\tau_\mathrm{FID}$ (assuming exponential decay), where $\tau_\mathrm{FID}$ is the free-induction decay time, so extending $\tau_\mathrm{FID}$ reduces the gate error rate.  Even if perfect spin echo pulses can be performed (on a time scale $t_g\ll \tau_\mathrm{FID}$), decay in the spin-echo envelope on a time scale $T_{2,\mathrm{echo}}$ will signal memory errors of typical size $\propto t/T_{2,\mathrm{echo}}$, where $t$ is the time elapsed since the beginning of a computation.  Extending $T_{2,\mathrm{echo}}$ therefore reduces the memory error rate.  While the historical approach has been to focus on spin-echo decay, in the context of quantum information processing it is necessary to consider both free-induction and spin-echo decay processes to eliminate both gate and memory errors.  For the reasons given above, these two processes must necessarily be considered independently, although they are both equally important aspects of the greater problem of coherent spin control.

\subsection{Free-induction decay}\label{sec:FID}

An electron spin in a magnetic field, confined to a semiconductor quantum dot or point defect in an s-type conduction band is well described by the Fermi contact Hamiltonian (Eq. (\ref{eq:FermiHamiltonianDot})) with the addition of an electron Zeeman term, on time scales short compared to the time at which the dipole-dipole Hamiltonian (Eq. (\ref{eq:Hdd})) becomes relevant (as noted in Sec. \ref{sec:DNPLimits}, dynamics under the dipolar Hamiltonian can be drastically suppressed in a number of cases due to Knight-field or quadrupolar inhomogeneity).  In an applied magnetic field, the Hamiltonian divides naturally into a secular part $H_0$ and a non-secular ``flip-flop'' term $V_\mathrm{ff}$:
\begin{equation}\label{eq:HFDivided}
 H_0=(b+h^z)S^z;\;V_\mathrm{ff}=\frac{1}{2}\left(h^+S^-+h^-S^+\right).
\end{equation}
Here, the electron Zeeman energy is $b=g^*\mu_\mathrm{B}B$ in an applied magnetic field $B$, and $\mathbf{h}=\sum_k A_k\mathbf{I}_k$ is the nuclear-spin field operator.  In the limit of very large $b$, we can consider evolution under $H\simeq H_0$ alone.  If the nuclear spin system is not in a specific eigenstate of the operator $h^z$, i.e., if the value of the nuclear field is unknown, the transverse spin will decay on a time scale $\tau_0\sim\sqrt{N}/A$. For a typical GaAs quantum dot containing $N\sim 10^5-10^6$ nuclei, this time scale is very short: $\tau_0\sim 1-10\,\mathrm{ns}$.

To extend the free-induction time, it is necessary to narrow the distribution of available values of $h^z$.  This can be done through dynamic polarization (see Sec. \ref{sec:DNP}), passive measurement \cite{Coish2004a,Klauser2006a,Stepanenko2006a,Giedke2006a,Barthel2009a}, or by actively driving the system toward a particular (known) state as in refs. \cite{Greilich2006a,Greilich2007a,Greilich2007b,Reilly2008b,Ribeiro2009a,Vink2009a}.

While polarization \emph{is} effective in reducing the spin-flip probability \cite{Burkard1999a}, it is relatively ineffective in extending the coherence time, resulting in a weak increase in the free-induction decay time for a polarization $p$ \cite{Coish2004a}: $\tau_\mathrm{FID}(p)=\tau_\mathrm{FID}(0)/\sqrt{1-p^2}$.  In this case, it may be advantageous to reduce the width of possible values for $h^z$ for $|p|<1$, as described above.  If this is done, the state of the nuclear spin system is said to be ``narrowed'' \cite{Klauser2006a}.  In the extreme case, where the nuclear spin system has been forced into an eigenstate of the operator $h^z$, $H_0$ will only induce simple precession of the electron spin, but decay can still occur due to $V_\mathrm{ff}$ or from internal dynamics in the nuclear spin system due to, e.g., dipolar coupling.

\begin{figure}[tb]
 \includegraphics[width=80mm]{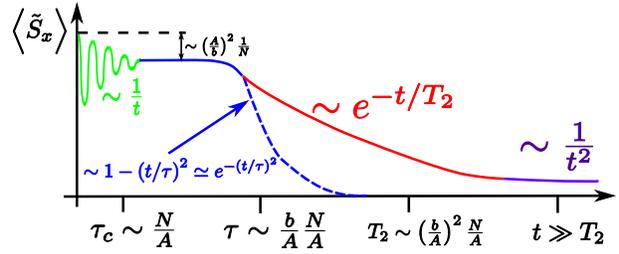}
\caption{\label{fig:fid} Illustration of the free-induction decay for the transverse components of a central spin in the rotating frame.  The spin is coupled to a bath of $\sim N$ nuclear spins via the contact interaction (Eq. (\ref{eq:FermiHamiltonianDot})), assuming an initial ``narrowed'' distribution for the nuclear field.  The sketch is accurate when the nuclear dipole-dipole interaction (Eq. (\ref{eq:Hdd})) is negligible and when the electron Zeeman splitting is large compared to the hyperfine coupling strength ($b\gtrsim A$). The power laws shown here at short and long times apply to an electron in a two-dimensional quantum dot.  See the text for a discussion of the various stages of decay.}
\end{figure}

On time scales where the dipolar coupling can be ignored, the problem of free-induction decay for a narrowed nuclear spin state has been investigated in great detail (see Fig. \ref{fig:fid} for an illustration of the electron-spin decay in a large magnetic field $b\gtrsim A$, where most perturbative theories can be controlled).  There is a small partial power-law decay on a time scale $\tau_c\sim N/A$, where $N$ is the typical number of nuclear spins with appreciable coupling constants $A_k$ \cite{Khaetskii2002a,Khaetskii2003a,Coish2004a} (green curve in Fig. \ref{fig:fid}), followed by a quadratic shoulder \cite{Yao2006a,Liu2007a} (blue curve in Fig. \ref{fig:fid}), which becomes exponential in the Markovian regime, typically for $b\gtrsim A$ \cite{Coish2008a} (red curve in Fig. \ref{fig:fid}), and decays to zero with a long-time power-law tail \cite{Deng2006a,Deng2008a} (violet curve in Fig. \ref{fig:fid}).  In the Markovian regime $b\gtrsim A$, the majority of the decay will be close to exponential, due to the difference in free-induction decay time and bath correlation time: $\tau_c\sim N/A < T_2\sim (b/A)^2 N/A$ \cite{Coish2008a}.

It is important to note that Fig. \ref{fig:fid} focuses on the free-induction decay for an electron in a two-dimensional quantum dot.  Many features of this sketch are non-universal, depending on the shape and dimensionality of the electron wave function.  In particular, one-dimensional quantum dots, such as those realized in carbon nanotubes \cite{Churchill2009a,Churchill2009b} should show a comparatively much faster decay for similar coupling strength, and may not admit an exponentially-decaying solution \cite{Coish2008a}.

The low-field regime ($b<A$) can be explored in a controlled way where exact solutions are available.  Specifically, in the case of a fully polarized nuclear spin system \cite{Khaetskii2002a,Khaetskii2003a}, for uniform coupling constants $A_k=A_0$ \cite{Coish2007b,Zhang2006a,Bortz2007a}, with exact numerical diagonalization of small systems \cite{Schliemann2002a,Schliemann2009a}, or from Bethe Ansatz solutions \cite{Bortz2007b}.  Alternatively, new work suggests that a resummation technique may allow for a controlled perturbative calculation of electron spin dynamics even at relatively low magnetic fields \cite{Cywinski2009a,Cywinski2009b}.

\subsection{Spin-echo}
Spin echoes were first investigated by Hahn \cite{Hahn1950a}, who showed that some of the coherence lost during free evolution of spins could be recovered with the application of an appropriate rf pulse.  A phenomenological theory of spin-echo decay for spins interacting with a spin environment was developed, initially by Herzog and Hahn \cite{Herzog1956a}, based on work by Anderson and Weiss on linewidth narrowing \cite{Anderson1953a}.  This theory, known as ``spectral diffusion'' assumes that the energy splitting of a central spin results from its interaction with other environmental spins.  These environmental spins undergo temporal fluctuations dictated by the dipole-dipole Hamiltonian (Eq. (\ref{eq:Hdd})), resulting in a randomized precession frequency for the central spin, and consequent decay.  While the earliest theories of spectral diffusion assumed Gaussian diffusion of the central-spin precession frequency, resulting in a decay envelope $\sim \exp\left[-(t/\tau)^3\right]$, subsequent theories emphasized the need to consider Lorentzian diffusion to recover typical experimentally observed decays closer to $\sim \exp\left[-(t/\tau)^2\right]$ \cite{Mims1961a,Klauder1962a}.

Interest in the spectral-diffusion problem has been rekindled in the last few years due to potential quantum-information-processing applications using spins in quantum dots \cite{Loss1998a,Cerletti2005a}, phosphorus donors \cite{Kane1998a}, NV centers in diamond \cite{Wrachtrup2006a}, and molecular magnets \cite{Leuenberger2001a,Lehmann2007a,Ardavan2007a}.  De Sousa and Das Sarma revisited the spectral diffusion problem, introducing stochastic flip-flops due to dipolar coupling \cite{DeSousa2003a,DeSousa2003b}, giving rise to a decay of the form $\sim\exp{\left[-(t/\tau)^3\right]}$.  Later, more microscopic descriptions have been given \cite{Witzel2005a,Witzel2006a,Yao2006a,Saikin2007a}, which show decay envelopes closer to gaussian $\sim\exp{\left[-(t/\tau)^2\right]}$, in agreement with experiments \cite{Klauder1962a,Abe2004a}.  However, these theories are valid only at very large magnetic fields, where the electron-nuclear flip-flop term ($V_\mathrm{ff}$ in Eq. (\ref{eq:HFDivided})) can be neglected or included perturbatively.  New work by Cywinski \emph{et al.} may solve this problem \cite{Cywinski2009a,Cywinski2009b} with a resummation of the most relevant terms, but is still limited to short times in the limit of large magnetic field.  Although these authors typically cite applications for single electron spins in quantum dots or bound to donor impurity sites, the same general theory has also been applied to decay of spin coherence in molecular magnets \cite{Troiani2008a} and to nitrogen vacancy (NV) center spins in diamond \cite{Maze2008b}. 

Experiments on single-spin echoes have been performed in the singlet-triplet subspace of a two-electron gated double quantum dot \cite{Petta2005a} and for single electrons in a double dot \cite{Koppens2008a}.  These studies tend to be limited to relatively low magnetic fields to limit the electron-spin-resonance (ESR) excitation frequency and consequently, the effects of photon-assisted tunneling \cite{Kouwenhoven1994a}.  New methods for single-spin rotation may be necessary to allow fast pulses at high magnetic fields.  These methods include those based on the spin-orbit interaction \cite{Nowack2007a}, nuclear Overhauser field gradient \cite{Laird2007a,Rashba2008a}, motion of the quantum dots in an applied magnetic field gradient \cite{Tokura2006a,Pioro-Ladriere2008a}, or the exchange interaction \cite{Coish2007a}.  In self-assembled quantum dots, a wide range of exciting new optical techniques for single-spin control have been developed over the last 2-3 years \cite{Li2003a,Berezovsky2008a,Xu2008a,Press2008a,Greilich2009a}.  Some of these same methods have been demonstrated for NV centers in diamond \cite{Santori2006a}, showing promise for extremely fast spin manipulation.

In addition to spin-echo envelope \emph{decay}, electron spin-echo envelope modulation (ESEEM) \cite{Rowan1965a,Mims1972b} is often observed.  ESEEM signals the presence of the anisotropic hyperfine interaction (Sec. \ref{sec:ahf}), allowing, in principle, for universal control of the nuclear spins through control of the electron transitions \cite{Hodges2008a}.  ESEEM introduces an additional modulation for electrons bound to phosphorus donor impurities in silicon \cite{Witzel2007d} due to anisotropic hyperfine interaction from $sp$-hybridized electron states, and has been analyzed for NV centers in diamond \cite{Childress2006a}.

\subsection{Multi-pulse and dynamical decoupling}\label{sec:dd}
A more powerful method of coherence preservation than the conventional (Hahn) spin echo is dynamical decoupling, which typically consists of a train of many pulses designed to suppress more general forms of decoherence.  For a general review of dynamical decoupling methods, see the book by Haeberlen \cite{Haeberlen1976a}.  Multi-pulse sequences have been investigated in several papers in connection with nuclear-spin induced decoherence \cite{Yao2007a,Witzel2007a,Zhang2007a,Zhang2008a}.  While earlier work on dynamical decoupling relied on a time-periodic sequence of pulses to remove evolution from an unwanted part of the Hamiltonian, more recently \emph{concatenated} decoupling schemes have been introduced \cite{Khodjasteh2005a}, which have a recursive structure, and can therefore eliminate a larger class of errors.  Concatenated schemes have been applied to the problem of nuclear-spin-induced decoherence \cite{Yao2007a,Witzel2007b}.  Recently, a new optimal set of pulses have been developed and applied to a related quantum decoherence model (the spin-boson model) \cite{Uhrig2007a}, which was later shown to be universally applicable to an arbitrary dephasing Hamiltonian, and applied to the problem of electron-spin decoherence in a nuclear spin bath (Lee et al. \cite{Lee2008a}).

New techniques, for example, employing an Euler-Lagrange equation for maximizing fidelity \cite{Gordon2008a} may lead to further improvements, and recent work \cite{Khodjasteh2009a} suggests that quantum error correction can be performed 'in line' using dynamical decoupling pulses.

Closely related to dynamical decoupling is the idea that spin coherence in a nuclear spin bath can be extended with continuous resonant excitation.  Recent experimental and theoretical work has shown that driven Rabi oscillations decay slowly (according to a power law) and at a long time scale under resonant excitation in quantum dots \cite{Koppens2007a} and NV centers in diamond \cite{Hanson2008a} in a static nuclear field.  Quantum corrections to this problem have been calculated \cite{Rashba2008a}, and  decay in the presence of dipolar interactions has been investigated \cite{Dobrovitski2009a}.

\section{Conclusions and outlook}\label{sec:conclusions}

We have given an overview of the physics of nuclear spins in nanostructures. The systems of interest include quantum dots, donor impurities, nanotubes, NV centers in diamond, and molecular magnets, where interaction with localized electrons plays a crucial role.  Our focus was on two main aspects that have been at the focus of recent studies: nuclear spin polarization and electron-spin decoherence in the presence of a nuclear environment.

There are a number of pressing issues related to the manipulation of nuclear magnetism in nanostructures, and the extension of single-spin coherence times in the presence of a nuclear spin environment.  Among the most important questions are: What will be the role of ``imperfect'' (finite-bandwidth) pulses in dynamical decoupling experiments?; How strong and fast can single-spin rotations be performed (in particular, which of the methods discussed in Sec. \ref{sec:dd} will allow the highest level of control)?; Will it be possible to substantially further narrow the nuclear-field distribution in single gated quantum dots, approaching the level that has been achieved optically in ensembles of self-assembled dots \cite{Greilich2008a}?; and Will it be possible to engineer diffusion barriers to control spin diffusion and preserve a local nuclear Overhauser field?

We believe that many of these questions will be answered in the next 2-3 years, but reaching a complete theoretical understanding of the underlying phenomena as well as designing and executing relevant experiments will be a significant challenge.

\begin{acknowledgement}
We acknowledge funding from NSERC, QuantumWorks, an Ontario PDF (WAC), and a CIFAR Junior Fellowship (WAC).
\end{acknowledgement}

%
\bibliographystyle{pss}
\bibliography{pssb.200945229}
%
%
%
%
%
%
%
%
\end{document}